\documentclass[11pt]{article} 
\usepackage{amsmath,amsfonts,amssymb,mathtools}
\usepackage{xspace} 
\usepackage{pifont} 
\usepackage{color}
\usepackage{slashed} 
\usepackage{bm} 
\usepackage{graphicx}
\usepackage{hyperref}
\usepackage[utf8]{inputenc}
\usepackage{physics}

\DeclareMathOperator{\arcosh}{arcosh}
\newcommand{\LQCD}{\Lambda_\mathrm{QCD}}
\newcommand{\tcut}{t_\mathrm{cut}}

\numberwithin{equation}{section}
\usepackage{authblk}

\date{}                     

\title{The Time Substructure of Jets and Boosted Object Tagging}
\author[ ]{Matthew D. Klimek}
\affil[ ]{Laboratory for Elementary Particle Physics, Cornell University, Ithaca, NY, 14853, USA\protect\\
Department of Physics, Korea University, Seoul 02841, Republic of Korea}

\begin{document}
\maketitle


\begin{abstract}
We initiate the study of the time substructure of jets, motivated by the fact that the next generation of detectors at particle colliders will resolve the time scale over which jet constituents arrive.
This effect is directly related to the fragmentation and hadronization process, which transforms partons into massive hadrons with a distribution of velocities.
We review the basic predictions for the velocity distribution of jet hadrons, and suggest an application for this information in the context of boosted object tagging.
By noting that the velocity distribution is determined by the properties of the color string which ends on the parton that initiates the jet, we observe that jets originating from boosted color singlets, such as Standard Model electroweak bosons, will exhibit velocity distributions that are boosted relative to QCD jets of similar jet energy.
We find that by performing a simple cut on the corresponding distribution of charged hadron arrival times at the detector, we can discriminate against QCD jets that would otherwise give a false positive under a traditional spatial substructure based boosted object tagger.
\end{abstract}

\section{Introduction}

Of the various objects that are reconstructed by particle collider experiments, jets are unique in that they are collections of particles.
Because individual jet constituents have different velocities, they arrive at the detector over some finite span of time.
On dimensional grounds, we can estimate that the typical scale of the Lorentz boost of a jet constituent is $\gamma = E/m \sim E_j/nm_h$, where $E_j$ is the jet energy, $n$ is the hadron multiplicity of the jet, and $m_h$ is a hadron mass.
The corresponding scale of the spread in arrival times at a detector a distance $R$ from the interaction point is then of order $\delta t\sim R\,\delta v\sim R\gamma^{-2}$. 
To estimate an upper bound on this time scale, we can take $m_h$ to be the mass of heaviest hadron that is typically expected to be present in the jet, which for $E_j\gtrsim m_Z$ is the proton \cite{pdg}.
For $R\sim$~1 m, $E_j\sim$~100 GeV, $n\sim$~10, and $m_h~\sim$~1 GeV, we have $\delta t\sim$~30 ps.

This duration is shorter than current calorimeter-based timing resolution, which is limited to at best about 150 ps for particle energies greater than 50 GeV \cite{delRe:2015hla}.
However, it serendipitously coincides with the resolution expected to be achieved by a new timing detector to be installed as part of the Compact Muon Solenoid (CMS) detector for the High Luminosity phase of the Large Hadron Collider (HL-LHC) \cite{Collaboration:2296612}.
The main motivation for this design is to cope with increased levels of pileup.
Instantaneous luminosity at the HL-LHC will go up by a factor of about 4 compared to the value at Run II of the LHC.
This level of timing precision allows individual bunch crossings to be resolved.
By effectively dividing each bunch crossing into separate snapshots, pileup can be reduced.

Presented with a new technical capability, we are in a position to ask what new physical phenomena it will reveal or searches it will enable, beyond its design purpose.
Indeed, the application of high resolution timing in the context of searches for long lived particles has already been suggested \cite{Liu:2018wte,Cerri:2018rkm}.
It will also allow, for the first time, the intrinsic time structure of jets to be routinely recorded.
In this work, we take the first steps toward describing the structure that will be accessible to this and future generations of timing detectors and explore one potential application: tagging of boosted objects.

The spread in arrival time is due to the fragmentation and hadronization process through which (nearly) massless partons are packaged into massive hadrons which may have velocities significantly less than one.
Given the non-perturbative nature of hadronization, it is currently unknown how to calculate these effects from first principles.
We must therefore resort to phenomenological models.
Fortunately, physically well motivated and phenomenologically successful models are known, and indeed form the basis of the Monte Carlo studies that have enabled collider discoveries to date.

These models present a clear picture of jets which generically contain temporal structure accessible to the CMS HL-LHC timing detector or similar detectors at future experiments.
These predictions may be derived in the rest frame of the system that creates the partons that in turn manifest as jets.
If this system is itself boosted, as in the case of a highly boosted hadronically decaying resonance, the corresponding velocity spectrum of jet constituents will be modified from the rest frame prediction.
We will demonstrate that this information can be harnessed via the timing detector and used as a boosted object tagger.

The remainder of this work is organized as follows.
In section \ref{sec:hadr}, we derive generic predictions for the velocity spectrum of jets based on a model of hadronization.
In section \ref{sec:boosted}, we apply this information to the problem of tagging boosted hadronically decaying objects.
Finally, we conclude in section \ref{sec:concl}.

\section{Jet time structure from hadronization}\label{sec:hadr}

\begin{figure}
	\begin{center}
		\includegraphics[width=0.4 \linewidth]{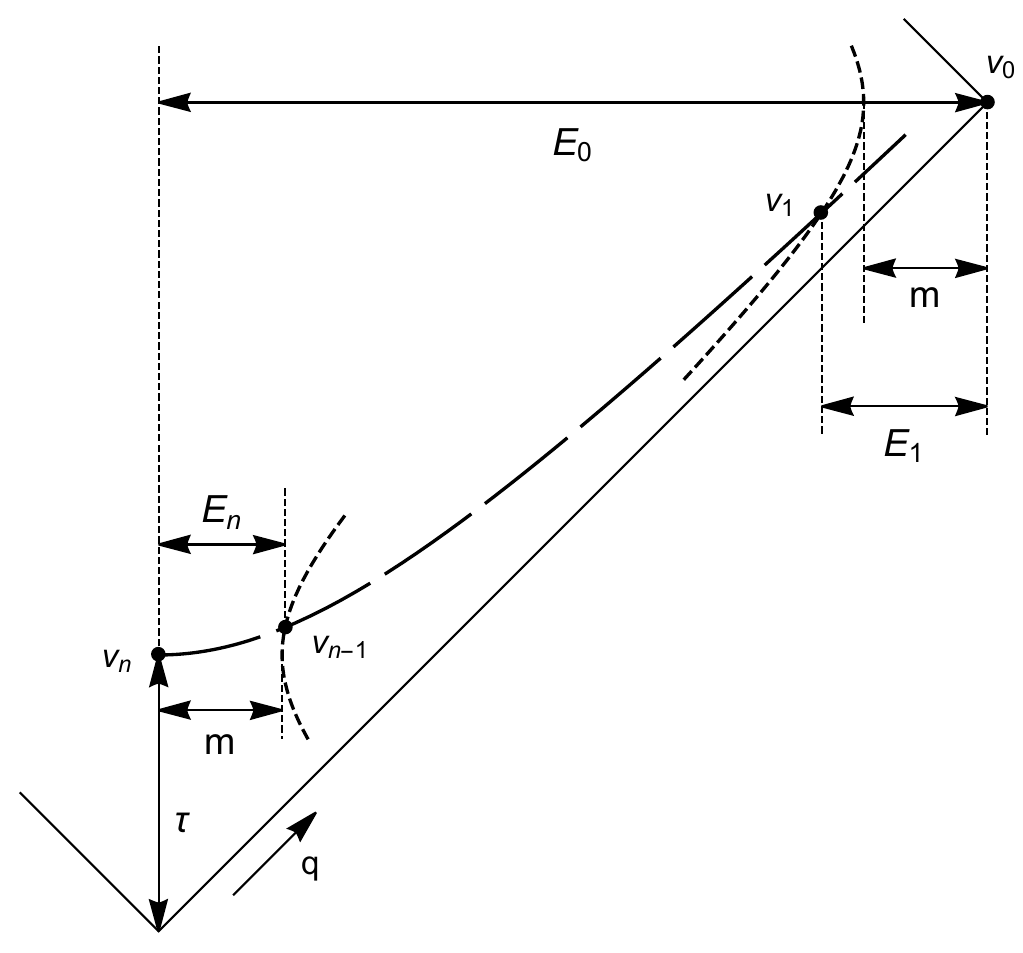}
	\end{center}
	\caption{Illustration of the features of the Lund string model described in the text. All distances in this figure are given by the labeled quantity divided by the string tension $\kappa$.}
\label{fig:lund} 
\end{figure}

Let us review the predictions of one hadronization model, the Lund string model \cite{Andersson:1983ia}, which forms the basis of the hadronization component of the widely used PYTHIA package \cite{Sjostrand:2006za, Sjostrand:2014zea}.
For the sake of simplicity, let us consider jets which are composed of only one type of hadron and which originate from massless quarks.
The model describes two quarks produced in a hard interaction as connected by a color string of constant tension $\kappa\approx 0.2$~GeV$^2$.
This results in a potential that grows linearly in the separation of the quarks, which is a manifestation of confinement.
Because the quarks are massless, they move on diagonal worldlines in a spacetime diagram out to a distance $\pm E_0/\kappa$, where $E_0$ is the initial energy of each quark, at which point all of the initial energy has been converted into string. They then turn around and move back toward their initial position.
As they travel apart, the string between them splits at a set of vertices, with new quark-antiquark pairs being formed from the vacuum at each vertex.
This process will be depicted in Figure \ref{fig:lund}.

Quarks from adjacent vertices are connected by fragments of the original string.
Such pairs are ultimately identified as the final hadrons.
In this picture, the mass of the hadron has a simple geometrical interpretation.
Quarks formed from the vacuum start with zero momentum, so if one imagines two neighboring vertices whose time coordinates are equal, the total energy, and thus the mass, of the quark-string system is then $m=\kappa \Delta x$, where $\Delta x$ is the spatial separation of the vertices. 
These two quarks move towards each other on diagonal worldlines, exchanging the energy of the string for momentum.
Their worldlines trace out a diamond on the spacetime diagram. 
The two quarks later meet again, exchanging places, before finally returning to the starting configuration.
The mass can then alternatively be expressed as $\sqrt 2 \kappa$ times the square root of the area of the diamond that the quarks will trace out.
Over one complete period, they trace out a total area $A$ equal to two diamonds, so we obtain the relation $m^2=\kappa^2 A$.
This relation holds in general, even when the two vertices are not simultaneous, as must be the case by Lorentz invariance.
It is straightforward to show that for general space and time separations $\Delta x$ and $\Delta t$, the mass of the system is then $m^2=\kappa^2[(\Delta x)^2-(\Delta t)^2]$.
This means that given the coordinates of one vertex, if we are to form a hadron of mass $m$, the neighboring vertex must lie on the hyperbola with radius $m/\kappa$ centered on the first vertex.

Because each vertex destroys the string in its future lightcone, all vertices have spacelike separations.
Vertices therefore cannot influence each other, and all vertices will tend to be found along the hyperbola of proper time $\tau^2 = t^2 - x^2$, where $\tau$ is the typical time for vertex formation.
This typical time is a parameter of the model which should be tuned by comparison with data.
However, QCD, whose dynamics are supposed to be captured by this model, has only one dimensionful parameter $\LQCD$, and so we expect $\tau\sim\LQCD^{-1}$.
This is indeed borne out by the data which require $\tau^{-1}\sim 250$~MeV.

Let us use this model to derive some approximate predictions for the general features of the velocity distribution of the hadrons.
Rather than working directly in terms of velocity, it will be convenient to use the rapidity $y=\frac{1}{2}\log((1+v)/(1-v))=\arcosh(E/m)$.
Let us consider the point $v_0$, which is the turn-around point for the original quark.
This is kinematically equivalent to a vertex where a quark is produced with zero energy. 
It has coordinates $v_0=(t_0,x_0)=(E_0/\kappa,E_0/\kappa)$.
This quark will combine with the antiquark formed at the next vertex $v_1=(t_1,x_1)$ to form a hadron of mass $m$.
That vertex must lie on the hyperbola given by $m^2 = (x_0-x_1)^2 - (t_0-t_1)^2$.
It is also likely to lie near the hyperbola of proper time $\tau$.
These two hyperbolae intersect at a single point which we will identify with $v_1$.
Solving for $x_1$, we can compute the separation of the vertices $x_0-x_1=E/\kappa$ from which we can estimate the rapidity of the fastest hadron in the jet.
If the initial energy of the original quark $E_0\gg\kappa\tau\sim1$~GeV, we find $x_1\approx E_0\kappa\tau^2/(m^2+\kappa^2\tau^2)$ which gives $y=\arcosh(\Delta x/m)\approx \log(2E_0 m/(m^2+\kappa^2\tau^2))$.
We see that the maximum rapidity varies logarithmically with the parton energy.
For $E_0=500$~GeV, we find $y\sim 5-6$ for a range of light hadron masses.
It is also clear geometrically from Figure \ref{fig:lund} that the leading hadron will have energy significantly greater than its mass.
The leading hadrons in the jet will be ultra-relativistic and travel to the detector at nearly the speed of light.

We see that the leading hadron has taken a fraction $z\sim(1+\kappa^2\tau^2/m^2)^{-1}$ of the original parton's energy.
For typical hadron masses this is $\lesssim10\%$.
The remaining energy will therefore still be large and this process will continue self-similarly.
After the $n$th hadron has split off, the remaining energy fraction will be approximately $(1-z)^n$.
Because the rapidity is proportional to the log of the energy, this implies that hadrons will have an approximately uniform distribution in rapidity.
This process will continue until there is insufficient energy to produce another hadron.
This final hadron arises from a vertex $v_n$ that falls near the bottom of the proper time hyperbola.
The remaining energy taken by this hadron is of order $\kappa\tau$.
We therefore obtain the relation $(1-z)^n\sim \kappa\tau/E_0$ which gives an estimate of the number of hadrons in the jet $n\sim z^{-1} \log(E_0/\kappa\tau)$.
We see that the jet multiplicity should also grow logarithmically with the energy.
We are also interested in the largest rapidity that the last hadron could have, as this will place a conservative bound on the arrival time of the trailing edge of the jet at the detector.
This bound will be saturated when the final vertex is at the very bottom of the hyperbola $v_n=(\tau,0)$.
We can again find the intersection of the proper time hyperbola with the mass constraint hyperbola to locate the preceding vertex $v_{n-1}$.
In this way we find the lowest rapidity $y_\mathrm{min}\sim\order{m^2/\kappa^2 \tau^2} \lesssim1$ independent of the parton energy.
We therefore expect that the jet will typically contain non-relativistic hadrons.

The CMS timing detector will be sensitive to charged particles and is expected to have a time resolution of about $\delta t\sim$~30 ps, with an acceptance limited by the requirement that a charged particle have sufficient $p_T$ to reach the detector in the 3.8 T magnetic field \cite{Collaboration:2296612}.
The leading hadrons travel at nearly the speed of light and mark the beginning of the jet.
Any hadron whose time of flight to the detector is within $\delta t$ of the light travel time will not be resolved separately from the beginning of the jet.
On the other hand, any hadron whose velocity is so low that it fails the acceptance threshold will not be detected.
This sets a range of rapidities from which information about the time profile of the jet may be extracted.
For a detector distance of 1 m from the interaction point, the maximum resolvable rapidity is 2.7 independent of the mass of the hadron.
The minimum detectable rapidity depends on the mass of the hadron, but is 2.3 for $\pi^\pm$, the lightest charged hadron.
We see from the preceding discussion that this range is always expected to be covered for relevant collider energies.
Because the rapidity distribution of hadrons is approximately flat, the number of time resolved hadrons in any given jet should have an approximately Poisson distribution.

PYTHIA8 \cite{Sjostrand:2014zea} implements a more complete version of the Lund model, and we use it to verify the predictions of the preceding paragraphs.
We use the default 4C tune \cite{Corke:2010yf} in which the model parameters have been adjusted to give good agreement with LHC data.
We generated samples of light quark jets at center of mass energies from 50 to 2000 GeV from $e^+e^-$ annihilation.
Jets are clustered using the Cambridge-Aachen (CA) \cite{Dokshitzer:1997in,Wobisch:1998wt} method with jet radius 0.4.
We select all jets with energy of at least 90\% of the parton energy.
Because we are working in the context of lepton collisions, we directly use angles, rather than  pseudorapidity, to compute particle distances.

\begin{figure}
	\centering
		\begin{minipage}{.45\textwidth}
		\centering
			\includegraphics[width=1 \linewidth]{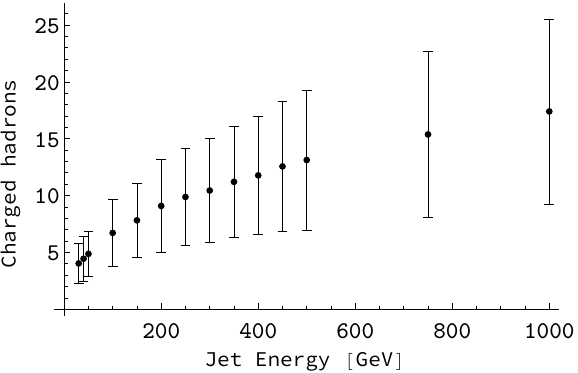}
		\end{minipage}\qquad\qquad
		\begin{minipage}{.45\textwidth}
		\centering
			\includegraphics[width=1 \linewidth]{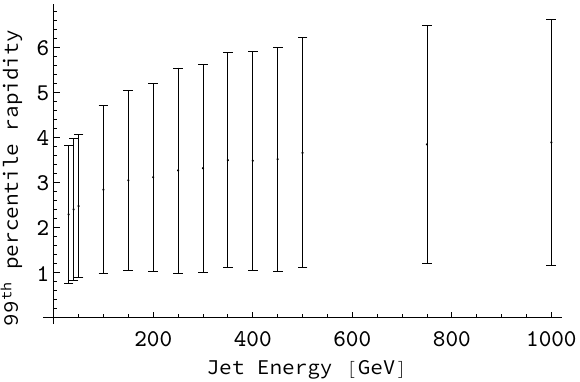}
		\end{minipage}
	\bigskip
		\begin{minipage}[t]{1\textwidth}
		\centering
			\caption{\emph{Left}: Jet charged hadron multiplicity mean and standard deviation, and \emph{Right}: 1st to 99th percentile rapidity range for light quark jets as a function of center of mass energy from PYTHIA8.}\label{fig:jetmult} 
		\end{minipage}
\end{figure}

\begin{figure}
	\begin{center}
		\includegraphics[width=0.5 \linewidth]{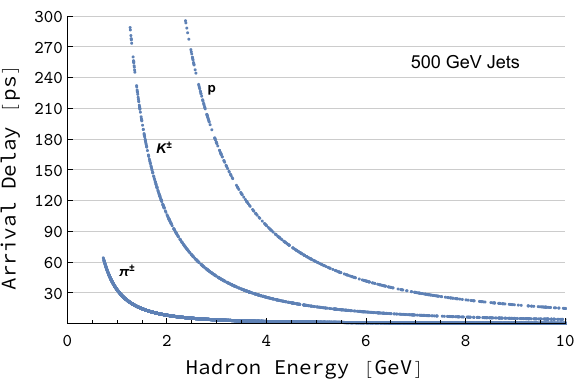}
	\end{center}
	\caption{Arrival time delay at 1 m versus energy for charged hadrons from an ensemble of 500 GeV jets. Horizontal lines are placed at intervals of the timing resolution of the CMS detector.}
\label{fig:timeenergy} 
\end{figure}

The results are compared to our predictions in Figure \ref{fig:jetmult}. 
We show the mean and standard deviation of the number of charged particles in each jet and confirm the approximately logarithmic growth in multiplicity with energy.
We also show the range of rapidities spanned by the hadrons in a sample of jets, with bars extending from the 1st to 99th percentile, and confirm that the highest rapidity varies logarithmically with parton energy while the lowest rapidity stays relatively constant and extends to values $\lesssim 1$.
In order to get a sense of the implications of these results in terms of hadron arrival time, in Figure \ref{fig:timeenergy} we show a scatter plot of arrival time delay at a distance of 1 m versus energy for charged hadrons in an ensemble of jets from 500 GeV light quarks.
Horizontal lines are drawn at intervals of the nominal CMS timing detector resolution.

These results demonstrate on general grounds that the velocity spectrum of jets will be directly measured by the next generation of collider detectors.
The details of that spectrum will constitute a novel probe of the hadronization process.
Improved understanding of hadronization is a primary goal of future electron-ion experiments \cite{Aschenauer:2019kzf,Aschenauer:2014cki}.
Measurement of the velocity spectrum provides an additional window into this process which has not yet been considered.

\section{Application to boosted objects}\label{sec:boosted}

Having demonstrated the basic characteristics that will determine the time profile of jets and noted that these time scales will be resolved by next generation detectors, let us consider one potential application to searches for new physics.
Many BSM models have signatures that contain boosted SM particles whose decay products are highly collimated \cite{Abdesselam:2010pt}.
Specialized techniques are needed to resolve them.
For recent reviews, see \cite{Marzani:2019hun,Larkoski:2017jix,Asquith:2018igt}.
In this section, we argue that the ability to resolve the time structure of jets can provide information on whether a jet is from a boosted object or not.

To motivate this, first consider the following two situations in the context of an electron-positron collider.
In the first case, we will imagine elastic $e^+ e^-$ scattering at some center of mass energy $\sqrt{s_\mathrm{el}}$.
In the second case, we imagine $ZZ$ production with one of the $Z$s decaying to electrons, but at a higher center of mass energy $\sqrt{s_{ZZ}}>\sqrt{s_\mathrm{el}}$. 
Let us choose $\sqrt{s_{ZZ}}$ to be sufficiently large so that for some kinematic configuration of the  $Z$ decay, one of the electrons has energy $\sqrt{s_\mathrm{el}}/2$ in the center of mass frame.
There is no difference between the two electrons in these two cases.
They will appear identically in any detector.\footnote{This is a leading order statement. Leptons that are produced at higher energies also have different probabilities of radiating extra photons or weak gauge bosons. However, due to the smallness of the relevant couplings, extra radiation is a subleading effect.}

Now consider the same two situations, but with $q\bar q$ production rather than elastic scattering and with hadronic $Z$ decay rather than leptonic.
We may again ask what is the difference between the two corresponding quarks.
At parton level they are indeed the same. 
The crucial difference in this case is that partons are not directly observable.
Rather, we observe the spray of hadrons that emerges from the showering and hadronization of the original parton.
The latter process may be viewed as the fragmentation of the color string of which the quarks are the endpoints.
In other words, the string, not the quark, is the observable object.
In the case of quark pair production, the string spans the two sides of the final state. Its energy is equal to the center of mass energy and it has no momentum.
In contrast, the string from the $Z$ decay is at rest in the $Z$ rest frame but boosted in the center of mass frame. Its mass, energy, and momentum are equal to the corresponding quantities of the $Z$.
These two objects are clearly different.
By Lorentz invariance, what determines the spectrum of the string fragmentation is the kinematics in the rest frame of the string, which was the point of view taken in deriving the predictions of section \ref{sec:hadr}.
If the string is boosted, the jet velocity spectrum will be shifted.

Let us consider a particular kinematic configuration of the $Z$ decays, where the decay axis is perpendicular to the $Z$ velocity in the $Z$ rest frame.
In the center of mass frame, the two jets will be pushed forward, giving a classic two-prong boosted jet substructure.
Such jets can be tagged using a recursive declustering method such as mass drop \cite{Butterworth:2008iy}.
In this technique, fat jets are first clustered with a large jet radius and then sequentially declustered until a splitting corresponding to a large mass, indicative of the hadronic decay of a heavy resonance, is found.
This is usually successful at removing QCD backgrounds, because QCD jets will tend to have smoother radiation patterns that do not contain a high mass splitting.
However, it is always possible that a QCD jet will by chance appear to have such a splitting, and this contributes to the fake tag rate.

Nevertheless, even when the radiation pattern of a QCD jet does fool the mass drop tagger, we should keep in mind that the string from which the radiation originated is very different from the string corresponding to a boosted object.
In the boosted $Z$ case, the two prong jet represents an entire string with mass $m_Z$, which has been boosted to energy $\sqrt s/2$.
The QCD jet, however, represents just one side of a string with mass $\sqrt s$.
This jet will contain a low velocity tail coming from the fragmentation of the center of the string.
In the $Z$ case, the low velocity tail has been boosted forward along with everything else.
This gives us additional information to cut on.
By cutting on the arrival time of the latest parton in the jet, we may be able to distinguish boosted jets from QCD fakes.

In order to demonstrate this, we produce two samples: hadronic $ZZ$ pairs and QCD multijets at $\sqrt s=$~500 GeV and 1 TeV from $e^+e^-$ in PYTHIA8.
These energies are chosen to correspond to those under consideration for the proposed International Linear Collider \cite{Behnke:2013xla}.
We select jets with $E_j>20$~GeV and cluster them with the CA algorithm with jet radius 1.2. 
We apply the mass drop tagger to both samples with parameters $\mu=0.5$ and $y=0.1$.
The results do not vary significantly under moderate changes of these parameters.
We select all jets that are tagged as coming from a boosted object with mass within 5 GeV of the $Z$ mass.
We find a QCD fake rate of about 1\%.
Although this fake rate is low, the multijet cross section is more than 20 times higher than the electroweak $ZZ$ production cross section, and so is not negligible.
We then cut on the arrival time of the latest charged hadron in the jet that is still above the threshold of the CMS timing layer, using a nominal distance of 1 meter from interaction point to timing detector.
Specifically, we keep only events whose latest detectable hadron arrives before some time $t_\mathrm{cut}$ after the earliest hadrons.
According to our preceding considerations, this should preferentially select jets from boosted objects whose low velocity tails have been boosted forward.

\begin{figure}
	\centering
		\begin{minipage}{.45\textwidth}
		\centering
			\includegraphics[width=1 \linewidth]{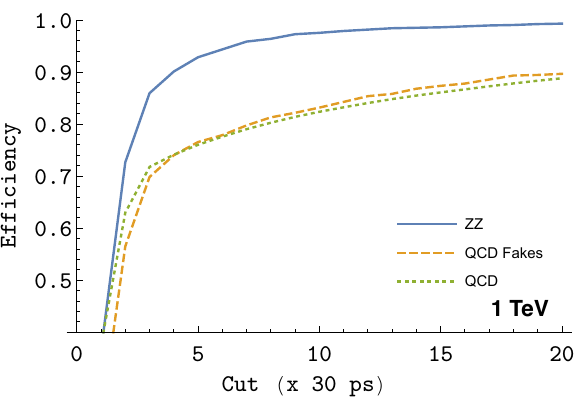}
		\end{minipage}\qquad\qquad
		\begin{minipage}{.45\textwidth}
		\centering
			\includegraphics[width=1 \linewidth]{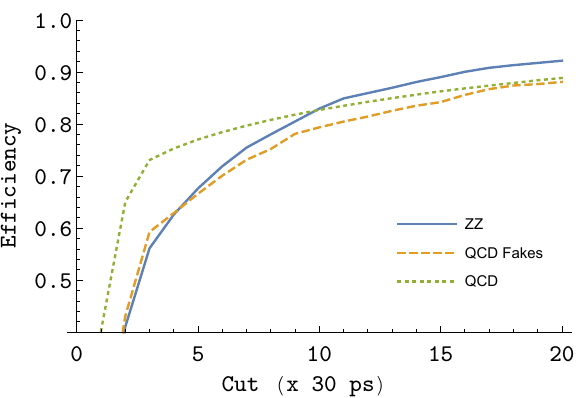}
		\end{minipage}
	\bigskip
		\begin{minipage}[t]{1\textwidth}
		\centering
	\caption{Selection efficiency as function of delayed arrival time cut on the slowest detectable charged hadron for hadronic $Z$ jets (blue, solid), QCD jets that fake $Z$ jets under mass drop tagging (orange, dashed), and all QCD jets (green, dotted) at 1 TeV (left) and 500 GeV (right).}\label{fig:efficiencies} 
		\end{minipage}
\end{figure}

The resulting selection efficiencies for boosted $Z$ jets, QCD fakes, and all QCD jets for a range of values of $\tcut$ are presented in Figure \ref{fig:efficiencies}.
We see that for values of $t_\mathrm{cut}$ of a few times the timing resolution, the efficiency for selecting boosted jets is about 20\% greater than for QCD jets at 1 TeV.
We emphasize that the timing cut performs equally well on generic QCD jets and those that have already managed to fake the mass drop tagger, and conclude that the timing based analysis is sensitive to different information in the event that is not utilized by the mass drop tagger.
Smaller but modest gains are also possible at 500 GeV.
Since the $Z$ decay products are less boosted, higher values of $\tcut$ are required.

\begin{figure}
	\begin{center}
		\includegraphics[width=0.5 \linewidth]{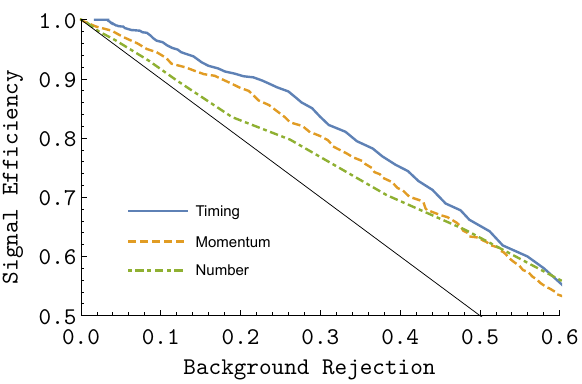}
	\end{center}
	\caption{ROC curves for the timing based cut (blue, solid), the momentum based cut (orange, dotted), and charged hadron number based cut (green, dot-dashed) at 1 TeV. The background is QCD jets that have already faked the Mass Drop cut.}
\label{fig:roc} 
\end{figure}

This analysis contains one parameter $\tcut$.
We characterize the performance of this cut with a Receiver Operating Characteristic (ROC) curve, in which we plot the efficiency of retaining the boosted events as a function of the fraction of background events that are rejected as $\tcut$ is varied.
So far we have been treating all particles as if they travel in a straight line to the detector.
In reality, a magnetic field is applied to the inner portions of the detector.
Charged particles move on curved paths in the magnetic field, with radius of curvature proportional to their transverse momentum.
This results in a path length to the detector which is longer for slower particles.
If a particle is sufficiently soft, its radius of curvature will be less than half the distance to the detector.
In this case, it will not hit the detector and is lost.
This sets the acceptance threshold for the detector, as was mentioned in the Introduction.
This effect also modifies the sensitivity to slow particles, as the time of flight now varies non-linearly with velocity.
In finding the ROC curve, we include the effect of a 3.8 T magnetic field for a timing detector at a distance of 1.2 meters from the interaction point, which corresponds to the CMS design.
The result is shown in Figure \ref{fig:roc}.

The timing-based cut proposed in this work makes use of velocity information, which is of course related to other kinematic information which is already measured by current technology. 
For example, this study relied on the slowest particles in a jet which will often be those with the lowest momentum.
One might expect that a simple cut on momentum would serve the same purpose.
Alternatively, since hadron multiplicity has a roughly logarithmic dependence on the rest frame energy of the quark-antiquark system, we could try to infer whether a jet came from a high mass string spanning the hard interaction, or from a boosted low mass string generated in a resonance decay by counting the number of hadrons in the jet.
In a sense, such objections are always fair.
Indeed, in any such tagging scenarios, we are trying to extract a single bit of information (tag or not) from a much richer set of information (momenta of all jet constituents, etc.).
In general, there will be a huge number of observables which have some overlap with the single bit that we are interested in.
We should therefore not expect to identify any single observable that tells us something that cannot be found in any form elsewhere.

On the other hand, there is reason to suspect that a direct measurement of velocity may be the most effective thing to do for an analysis such as the one suggested here.
Velocity is the natural quantity in which to talk about boosts.
Boosts preserve velocity ordering even among objects of different mass, whereas momentum ordering can change, and indeed jets are composed of particles with different masses.
Regarding hadron multiplicity, we note that despite the correlation of average particle number and hadronization energy, the scatter in the number of hadrons is very large.
Inspecting Figure \ref{fig:jetmult}, we see the number of particles from high energy jets is very often consistent with the number from a lower energy jet.
In addition, a two-prong jet from a boosted object decay captures both sides of the string, contributing a factor of roughly two in particle multiplicity compared to a hard QCD jet, partially compensating for the lower rest frame energy.
A cut on hadron number may therefore have only weak discriminating power.
To demonstrate these points, we perform the same ROC analysis as above by making a cut on lowest momentum or on charged hadron multiplicity.
The results are again shown in Figure \ref{fig:roc}, and we indeed see that their performance is not as good as the timing cut.

To clarify the difference between the timing and momentum cuts, we compare their performance at a fixed background rejection benchmark rate of 0.35.
For the timing cut, this requires $\tcut =660$~ps.
The results in Figure \ref{fig:roc} show that for the same background rejection rate, a momentum cut performs worse in signal efficiency.
To investigate why this is, we consider the momentum of all charged hadrons that arrive after $\tcut$ in the background sample of QCD fakes.
These are the particles that are providing the discriminating power.
Of these, 75\% are pions, with the others being kaons and (anti-)protons.
All of these pions have momentum less than 0.62 GeV, which is the momentum corresponding to a velocity that results in an arrival time equal to $\tcut$ at the pion mass.
Of the kaons and protons, 21\% have momentum less than 0.62 GeV.
The rest have momenta higher than the pions, but still arrive after $\tcut$ due to their greater mass.
As a result, a momentum cut would not be able to utilize these heavy hadrons while maintaining a similar cut on the pions which make up the majority of the hadrons.
Since the momentum cut will not be able to distinguish many of those QCD jets whose late hadrons are heavy, one would have to make the momentum cut stricter in order to achieve the same background rejection rate. 
However, this would result in a corresponding loss of signal efficiency, which is what we see illustrated in Figure \ref{fig:roc}.

We also note in passing that the distinction between momentum and velocity, which determines arrival time, is of course determined by the mass, or equivalently, the species of the hadron.
Comparison of arrival time measured by the timing detector and momentum measured by the tracker provides a means of identifying the species of the hadron, which information is valuable in many contexts, and we suggest this as a direction of further work.

\section{Conclusion}\label{sec:concl}

In this paper we have noted that the time structure of jets will be routinely observed at near future collider experiments.
We have elucidated the basic properties of their time profiles and suggested a possible use for this new information for the purpose of tagging boosted objects.
This is markedly different from the usual notion of jet substructure, which is concerned with the structure of the radiation that originates from the showering phase of parton evolution.
The time profile properties that we have studied come from the effects of hadronization.
Future measurements of these time profiles will also enable us to confront models of hadronization with data in a novel way.

We have demonstrated that a boosted object tagger incorporating timing information is complementary to traditional taggers, and can effectively discriminate against background events which may have faked other tests.
The timing information was also shown to be more effective than other related measurements.

This work was performed in the context of a lepton collider, which is simpler in a number of ways.
In particular, at a lepton collider there is no soft radiation from the underlying event.
Because the analysis we propose relies on the slowest, and therefore softest, particles in the event, it is particularly susceptible to contamination from soft background radiation.
We hope that pileup mitigation and jet grooming techniques that are necessary at high luminosity hadron colliders may still allow these techniques to be used in that context as well, but we leave this for future work.
Additionally, in a lepton collider context, the color strings in the final state do not connect to the initial state so it is easy to understand their properties.
Additional considerations would apply when trying to understand the time structure of jets from processes with colored initial states.

Given a new dimension of data about jets that will be available in the near future, we are sure that there is great potential for many other applications that are yet to be discovered. 
Additional work will also be needed to understand how to apply such techniques in the context of hadron colliders.
We suggest these as topics for future work.

\section*{Acknowledgements}
The author is grateful for enlightening conversations with Michelangelo Mangano, Maxim Perelstein, Matthew Schwartz, and Jesse Thaler, and acknowledges the support of the U.S. National Science Foundation, through grant PHY-1719877, the Samsung Science \& Technology Foundation, under Project Number SSTF-BA1601-07, and a Korea University Grant.

\bibliographystyle{utphys}
\bibliography{jettime}

\end{document}